\documentclass[epjCONF]{svjour}
\usepackage{graphicx}
\usepackage[varg]{txfonts} 
\usepackage[latin1]{inputenc}
\session-title{Fusion 11}
\begin{document}
\sloppy
\title{Static and Dynamic Chain Structures in the Mean-Field Theory}
\author{T.~Ichikawa\inst{1} \and N.~Itagaki\inst{1}
\and N.~Loebl \inst{2}
\and   J.~A.~Maruhn\inst{2,1}\thanks{Invited Speaker} 
\and V.~E.~Oberacker\inst{3} 
\and S.~Ohkubo \inst{4} \and B. Schuetrumpf\inst{2} 
\and A.~S.~Umar\inst{3}}

\institute{Yukawa Institute for Theoretical Physics, Kyoto University,
Kyoto 606-8502, Japan
\and Institut fuer Theoretische Physik, Goethe-Universitaet, 
Frankfurt am Main, Germany
\and Department of Physics and Astronomy, Vanderbilt University, 
Nashville, Tennessee 37235, USA
\and Department of Applied Science and Environment,
University of Kochi, Kochi 780-8515, Japan}
\abstract{
We give a brief overview of recent work examining the presence of
$\alpha$-clusters in light nuclei within the Skyrme-force Hartree-Fock
model. Of special significance are investigations into $\alpha$-chain
structures in carbon isotopes and $^{16}$O. Their stability and
possible role in fusion reactions are examined in static and
time-dependent Hartree-Fock calculations. We find a new type of shape
transition in collisions and a centrifugal stabilization of the
$4\alpha$ chain state in a limited range of angular momenta. No
stabilization is found for the $3\alpha$ chain.
} 
\maketitle
\section{Introduction}
\label{intro}
In light nuclei, cluster structures and large deformation may be
present even in the mean-field model \cite{clusters}.
For example, $^8$Be has an
$\alpha$-$\alpha$ structure of ``superdeformed'' nature, and the
degree of deformation is closely related to the orbits of valence
neutrons in neutron-rich Be isotopes \cite{Itagaki-Be,Ito}.  Also there
have been many discussions on the nature of the $0^+_2$ state of $^{12}$C
with well-developed $3\alpha$ structure, possibly a linear chain
\cite{Hoyle,Morinaga,Ikeda,Tohsaki}.
Another example is the $4\alpha$
linear chain band starting around the $4\alpha$ threshold energy region
in $^{16}$O suggested by Chevallier {\em et\ al.} \cite{Chevallier} through
the data analysis of the $^{12}{\rm C}+\alpha\to{}^8{\rm Be}+{}^8{\rm
Be}$ reaction.
This is supported by theoretical work\cite{Suzuki}; they
analyzed the decay widths and discussed that the observed states 
are possibly characterized by well-developed $4\alpha$ structure. 

Concerning the stability of the linear chain states, in conventional
models, it has been usually studied through the analyses of small
vibration around the equilibrium configuration \cite{Horiuchi}.
However, the bending motion is the most important path for the
transition to the low-lying states \cite{Umar}.  Therefore, it is
necessary to calculate the stability in a wide wave function space,
which covers not only the linear chain configurations but also the
lower excited states.  Utilizing a rather large model space, we have
analyzed in neutron-rich C isotopes that adding valence neutrons
increases the stability of the linear chain states with respect to the
bending motion \cite{Itagaki-C,Maruhn-C}. The basic instability still
remains, though. 

Even if they are fundamentally unstable, the chain structures could
still appear as intermediary configurations in a collision involving
$^8$Be and/or $\alpha$-particles, and one of the aims of this paper is
to explore this possibility using time-dependent Hartree-Fock (TDHF) methods.
It is generally acknowledged that the TDHF theory provides a
useful foundation for a fully microscopic many-body theory of
low-energy heavy-ion reactions \cite{Ne82}. Earlier TDHF calculations
of collisions between light nuclei involving cluster-like
configurations have been made for the study of nuclear molecular
resonances~\cite{US85}. However, due to the lack of computational
resources these calculations suffered from numerical imprecision as
well as unphysical symmetry assumptions, such as collisions being
restricted to axial symmetry.  Current TDHF calculations are performed
with high numerical precision and with no symmetry assumptions as well
as using modern Skyrme forces.  Recently, we have shown that when TDHF
is combined with the density-constraint method~\cite{CR85} dynamical
potentials and ion-ion interaction barriers can be accurately
reproduced~\cite{UO06b,UO08a,UO09b}.  Thus an alternative approach
would be to investigate the formation and stability of the linear
chain configuration using the fully microscopic and dynamical TDHF
theory.

An important mechanism
to stabilize the linear chain state could be the rotation of the
system; a large moment of inertia such as in the linear-chain configuration
is favored if large angular momentum is given to the nucleus.
If the angular momentum is too high, however, it leads to the fission of the
nucleus, and there must exist a region of angular momentum where the
linear chain configuration is stabilized.  We focus on this mechanism
and discuss the angular momentum of the linear-chain configuration
based on cranked Hartree-Fock.

\section{The $3\alpha$ chain state in nuclear  collisions}
In this work we study TDHF collisions which reproduce the linear chain
configuration and subsequently decay to lower-energy configurations of
the system. To our knowledge such mode changes have never been
observed in TDHF calculations previously and appear to simulate the
suggested astrophysical mechanism for the formation of $^{12}$C
nuclei.
\begin{figure}[!htb]
\begin{center}
\includegraphics*[scale=0.30]{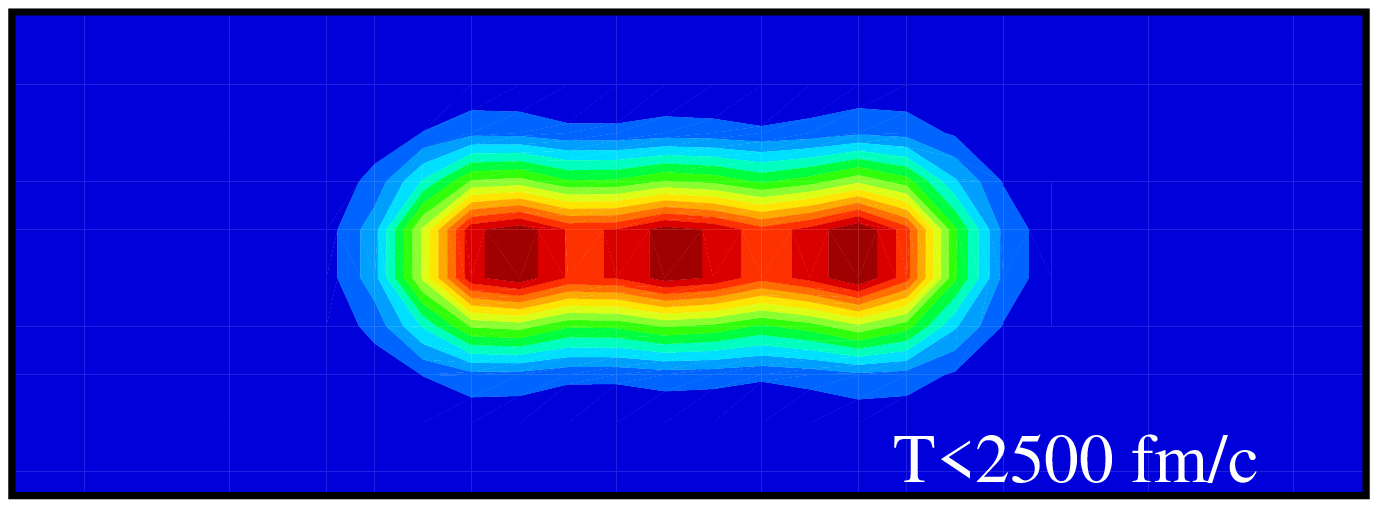}\\ \vspace{-0.03in}
\includegraphics*[scale=0.30]{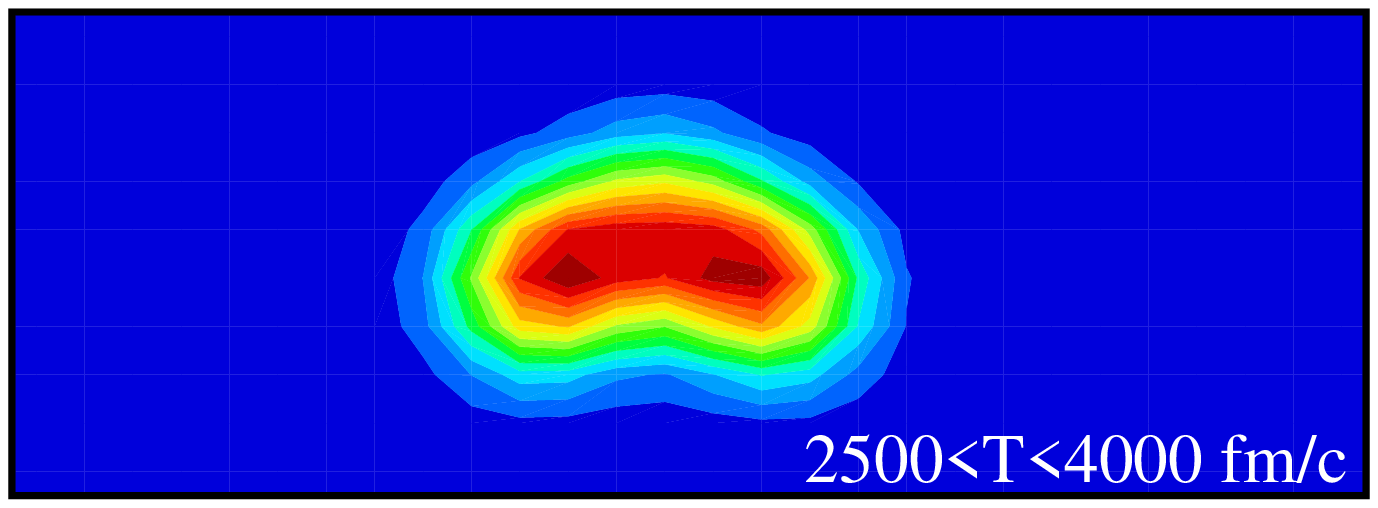}\\ \vspace{-0.03in}
\includegraphics*[scale=0.30]{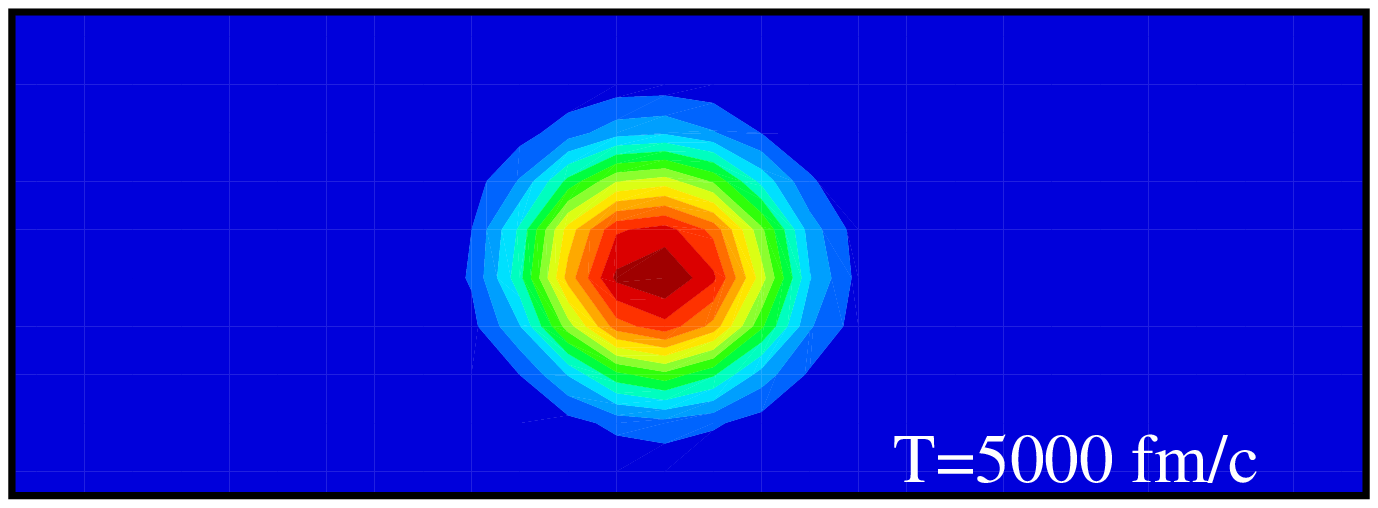}
\end{center}
\caption{\label{fig1}Selected density profiles from TDHF time-evolution
of the $^{4}$He+$^{8}$Be head-on collision for initial
Be orientation angle $\beta=0^{\circ}$ using the SLy4 interaction.
The initial energy is $E_{\mathrm{c.m.}}=2$~MeV.}
\end{figure}

In our numerical calculations we have chosen a Cartesian box which is
$40$~fm along the collision axis and $24$~fm in the other two
directions.  Calculations are done in 3-D geometry and using the full
Skyrme force (SLy4)~\cite{CB98} as described in Ref.~\cite{UO06}.
Using different modern parametrizations of the 
Skyrme force all show the same phenomena.  We have chosen to study
two different collisions leading to the linear chain configuration,
first the $^{4}$He+$^{8}$Be system and then the triple collision of
three $^{4}$He nuclei, which may be astrophysically much less
probable. The Hartree-Fock (HF) state for the $^{8}$Be nucleus is
axially symmetric. In Fig.~\ref{fig1} we show three snapshots from the
long time evolution of the $^{4}$He+$^{8}$Be collision at zero impact
parameter and $E_{\mathrm{c.m.}}=2$~MeV. The top panel of
Fig.~\ref{fig1} shows the linear chain configuration about which the
system oscillates for times less than $2,500$~fm/$c$.  In particular,
it is remarkable that the moving clusters do not equilibrate while
moving inside the linear chain state but retain their $2p-2n$
character, where one observes a complex quasiperiodic motion with
little damping up to this time.  Around $2500$~fm/$c$ the system
starts to bend and acquires a somewhat triangular shape as shown in
the middle panel of Fig.~\ref{fig1}.  The system still retains its
cluster character with the center cluster moving off the reaction
plane cut shown in the figure, but can be clearly observed in
volumetric three-dimensional movies of the collision process.  The
bending motion, where the center cluster oscillates somewhat
perpendicular to the left and right clusters continues for
approximately $1000$~fm/$c$, with very little damping.  Finally, at
even longer times the system relaxes into a relatively more compact
shape (bottom panel of Fig.~\ref{fig1}.).  Such mode changes, where
the dynamical energy in the longitudinal direction is converted to a
transverse mode, while the system retains its cluster structure have
never been seen in previous TDHF calculations albeit this would not
have been possible in calculations imposing axial symmetry.  Even in
three-dimensional calculations, for an exactly central collision, the
axial symmetry would be preserved under ideal theoretical and
numerical conditions. Therefore the meaning of head-on collision
($b=0$) should be interpreted to have a small dispersion around this
value, which facilitates the mode change even for exactly central
collisions.
\begin{figure}[!htb]
\begin{center}
\includegraphics*[scale=0.3]{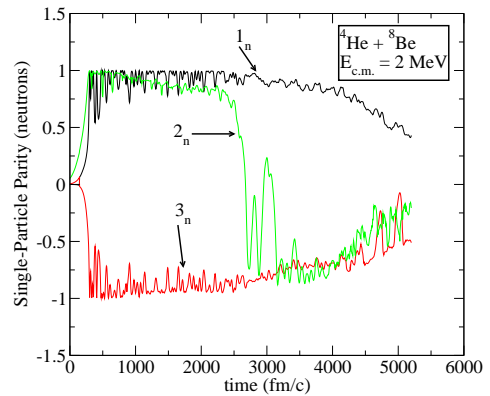}
\end{center}
\caption{\label{fig2a} Single-particle parities of the neutron states during the
collision of the $^{4}$He+$^{8}$Be system as a function of time at $E_{\mathrm{c.m.}}=2$~MeV.}
\end{figure}

We were also successful, for the first time, in creating a static
Hartree-Fock linear chain state orthogonal to the ground state in a
fully three-dimensional geometry.  This was achieved by initializing
one of the single-particle states to be in the $s-d$ shell with
positive parity rather than in the $p$ state with negative parity.
This results in a linear chain state similar to the one shown in the
top panel of Fig.~\ref{fig1}. The fact that this state has two
positive parity and one negative parity state proves that it is
exactly orthogonal to the $^{12}$C ground state which has one positive
parity and two negative parity states.  A similar dependence on parity
was also studied in cluster model calculations~\cite{IO06}.  In order
to relate this observation to the dynamical mode changes discussed
above we have used the density constraint method to calculate the
potential energy and the single-particle parities during the TDHF
time-evolution.  In Fig.~\ref{fig2a} we show the neutron
single-particle parities as a function of collision time.  What is
striking is that the combined system initially has the same parity
signature as the static linear chain state but at the time of bending
one of the positive parity states starts to decay towards negative
parity and this decay continues as the system becomes closer to the
parity signatures of the ground state. Oscillations in the numerically
calculated parities stem from the fact that these are done during the
dynamical evolution of the system.  The proton single-particle
parities are almost exactly the same as for the neutrons as
anticipated.
\begin{figure}[!htb]
\begin{center}
\includegraphics*[scale=0.3]{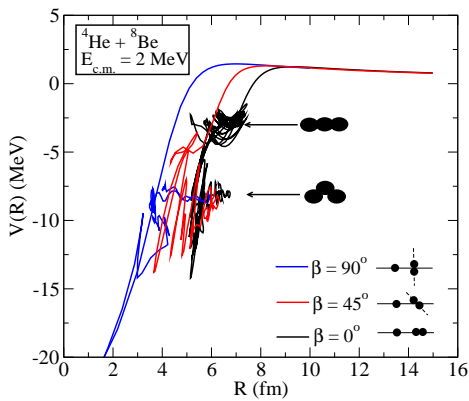}
\end{center}
\caption{\label{fig3a} Potential energy curves for the
  collision of the $^{4}$He+$^{8}$Be system as a function of $R$ for
  three initial alignments of the Be nucleus and at
  $E_{\mathrm{c.m.}}=2$~MeV.}
\end{figure}

In order to relate the observed mode changes more closely to the
intrinsic energy of the system we have also calculated the potential
energy of the system as a function of the ion-ion separation distance
$R$~\cite{UO06b}. For the calculation of $R$ we have used the hybrid
method described in~\cite{UO09b}, which relates $R$ to the quadrupole
moment of the system thus making it possible to have a consistent
definition of $R$ for large overlaps.  The calculated potential energy
curves are shown in Fig.~\ref{fig3a} as a function of three alignments
of the $^{8}$Be nucleus with respect to the collision axis labeled as
angle $\beta$, and for the entire duration of the collision process.
Since for the real system the angular momentum of $^{8}$Be is
projected all possible alignments of the $^{8}$Be nucleus needs to be
considered.  For all of the alignments the combined system climbs up a
shallow potential barrier height of approximately $1.24-1.44$~MeV, the
lowest barrier being that of the $\beta=0^{o}$ potential, thus making
this alignment most probable under threshold conditions.  For the
potential energy curve showing the head-on collision ($\beta=0^{o}$,
black curve) we observe that the system initially relaxes to a
relatively shallow metastable minimum and oscillates about this
minimum until approximately $T=2500$~fm/$c$ at which points it slips
down the curve towards the second configuration as indicated by three
alphas in a triangular configuration. After spending some time in this
configuration the system further slips down to even lower energy and
more compact configuration.  The potential energy curves corresponding
to $^{8}$Be initial alignment angles of $\beta=45^{o}$ and
$\beta=90^{o}$ (red and blue curves, respectively) undergo a different
behavior, bypassing the linear chain minimum but directly going to the
triangular and subsequently to the compact configuration, the
perpendicular energy collision attaining the most compact and lowest
energy configuration.  It is interesting to note that all of the
potential energy curves spend some time in the triangular
configuration.

\begin{figure}[!htb]
\begin{center}
\includegraphics*[scale=0.3]{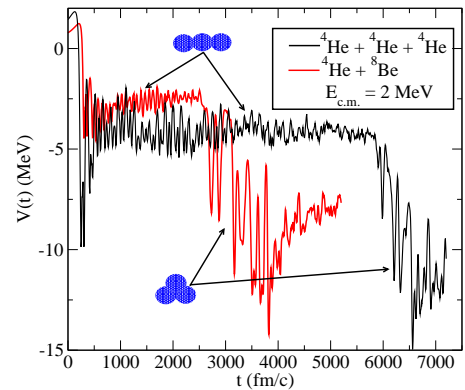}
\end{center}
\caption{\label{fig4a} Time development of the potential
  energy for the head-on collision of the $^{4}$He+$^{8}$Be and the
  $^{4}$He+$^{4}$He+$^{4}$He systems for $E_{\mathrm{c.m.}}=2$~MeV.}
\end{figure}

As an alternate collision leading to the same configuration we have
also studied the collision of three $^{4}$He nuclei, one at rest at
the origin of the collision axis and the other two on each side
boosted towards the center with $1$~MeV energy. In Fig.~\ref{fig4a} we
contrast the time dependence of the potential energies for the two
different collisions. We observe that the three $^{4}$He collision
process spends considerably longer time (about $6000$~fm/$c$)
undergoing quasiperiodic oscillations with very little damping in the
linear chain configuration before switching to bending and compact
modes.

\begin{figure}[!htb]
\begin{center}
\includegraphics*[scale=0.3]{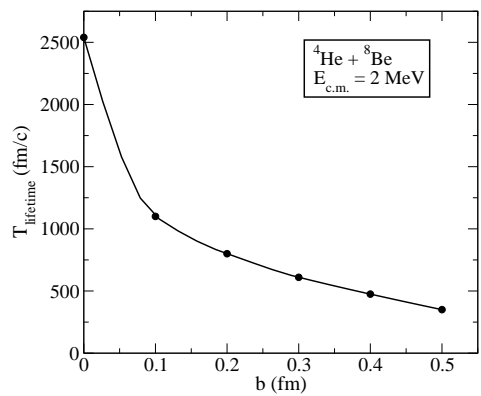}
\end{center}
\caption{\label{fig5a} Time spent in the linear chain configuration as
  a function of the impact parameter $b$ for the $^{4}$He+$^{8}$Be
  system at $E_{\mathrm{c.m.}}=2$~MeV and $\beta=0^{o}$ alignment.}
\end{figure}

In order to gauge the stability of the linear chain configuration we
have made systematic studies as a function of impact parameter and
center-of-mass energy, as well as a study using heavier Be isotopes to
determine the dependence on neutron number.  In Fig.~\ref{fig5a} we
show the dependence of the linear chain survival time on the impact
parameter for the $^{4}$He+$^{8}$Be system at
$E_{\mathrm{c.m.}}=2$~MeV and $\beta=0^{o}$. We observe that as the
impact parameter increases the survival time rapidly decreases.  This
decrease naturally happens slower (faster) for lower (higher)
energies. We have also done a similar study for the time spent in the
linear chain configuration as a function of the center-of-mass energy
for the $^{4}$He+$^{8}$Be system for $\beta=0^{o}$ alignment.  We
decreased the energy in steps of $0.1$~MeV to find the lowest energy
for which we form the linear chain configuration (at lower energies
the nuclei rebound due to Coulomb repulsion). At this energy of
$1.3$~MeV the lifetime of the linear chain configuration increases to
about $2875$~fm/$c$. As the energy is increased the lifetime decreases
gradually.  In order to study the dependence of the linear chain state
on the neutron number of the Be isotopes we have repeated all of the
above calculations using a $^{9}$Be nucleus instead. The calculations
were done by using all the time-odd terms in the Skyrme interaction
appropriate for an odd-A nucleus.  While we do find an analogous
behavior in this study, the lifetime of the linear chain state is
approximately $30$\% less than that of the corresponding $^{8}$Be
system.

\section{Cranked $4\alpha$ chain states}
To discuss the $4\alpha$ linear chain state in the rotational
frame, we perform cranked HF calculations. We self-consistently
calculate the cranked HF equation, given by $\delta\left<H-\omega
  J\right>=0$, where $H$ is the total Hamiltonian, $\omega$ is the
rotational frequency, and $J$ is the angular momentum around the $y$
axis.  We also perform TDHF calculations in order to discuss the
$4\alpha$ linear chain state in a collision situation.

We represent the single-particle wave functions on a Cartesian grid with
a grid spacing of 0.8~fm. The grid size is typically $24^3$ for ground
states and $32\times24^2$ for superdeformed states. This accuracy was
seen to be sufficient to provide converged configurations.  The
numerical procedure is the damped-gradient iteration method
\cite{gradient}, and all derivatives are calculated using the Fourier
transform method.

We take three different Skyrme forces: Sly6 as a
recent fit which includes information on isotopic trends and neutron
matter \cite{Cha97a}, and SkI3 as well as SkI4 as recent fits which
map the relativistic isovector structure of the spin-orbit force
\cite{Rei95a}.  SkI3 contains a fixed isovector part analogous to the
relativistic mean-field model, whereas SkI4 is adjusted allowing free
variation of the isovector spin-orbit term. Thus all forces differ
somewhat in their actual shell structure. Besides the effective mass,
the bulk parameters (equilibrium energy and density,
incompressibility, symmetry energy) are comparable.

\begin{figure*}[t]
\begin{center}
\includegraphics[width=0.6\linewidth]{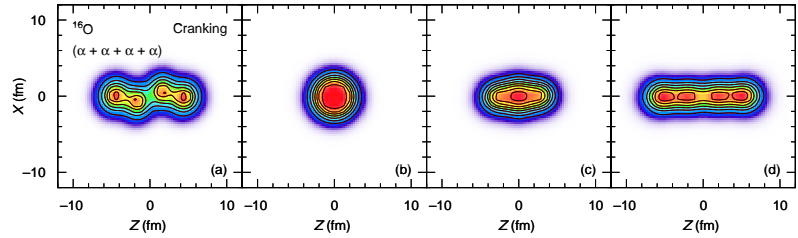}\\%
\end{center}
\caption{\label{den1} Total nucleon density distribution
 calculated using the cranking method for (a) the initial wave function,
 (b) the ground state, (c) the 
quasi-stable state, and (d) the $4\alpha$ linear chain state.  The
isolines correspond to multiples of 0.02 fm/c. We normalize the color to
the density distribution at the maximum of each plot.}
\end{figure*}

\begin{figure}[b]
\begin{center}
\includegraphics[width=0.6\linewidth]{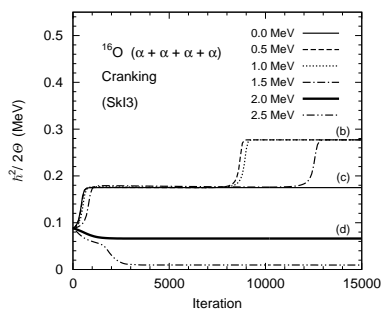}\\%
\end{center}
\caption{\label{ita1} Coefficient of the rotational energy,
 $\hbar^2/2\Theta$, calculated using the cranking method versus the HF
 iterations with various rotational  frequencies $\omega$. The symbols
 (b), (c), and (d) correspond to the 
 density distribution given in Fig.~\ref{den1}.}
\end{figure}

\begin{figure}[htbp]
\begin{center}
\includegraphics[width=0.6\linewidth]{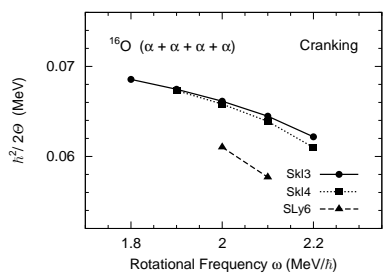}\\%
\end{center}
\caption{\label{iner1}Coefficient of the rotational energy,
 $\hbar^2/2\Theta$, calculated using the cranking method versus the
 rotational frequency $\omega$. The lines correspond to the different
 Skyrme forces as indicated.}
\end{figure}

\begin{figure}[h]
\begin{center}
\includegraphics[width=0.6\linewidth]{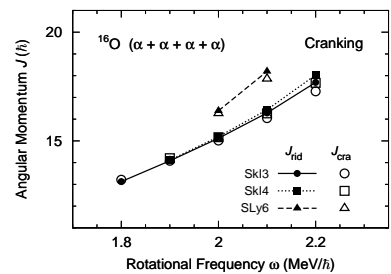}\\%
\end{center}
\caption{\label{mom1} Angular momentum calculated using the cranking
 method versus the rotational
 frequency. The lines with solid symbols denote the
 calculated results for the rigid-body moment of inertia, while the
 open symbols denote the results for the cranking method; both for the
 Skyrme forces as indicated.}
\end{figure}

We here discuss the stability of the $4\alpha$ linear chain
configuration in the rotating frame for $^{16}$O.  To this end, we
perform the cranked HF calculations with various rotational
frequencies, $\omega$. For the initial wave function, we 
use the twisted $4\alpha$ configuration, as shown in Fig.~\ref{den1}
(a). 
Note that this initial state violates axial symmetry,
which facilitates the transition of the initial state to low-lying states
including the ground state during the convergence process (this was
demonstrated for the carbon chain states in \cite{Itagaki-C,Maruhn-C}).
We calculate the rigid-body moment of inertia, $\Theta$, using the
total nucleon density at each iteration step.  We here only
consider rotation around the $y$ axis.

We first investigate the convergence behavior of the HF iterations.
To check this, we calculate the coefficient of the rotational energy,
given by $\hbar^2/2\Theta$, at each iteration step.  Figure~\ref{ita1}
shows the calculated results with various rotational frequencies
versus the iterations in the case of the SkI3 interaction.  The
initial state with the twisted linear chain configuration is not the
true ground state of the HF model space and the solution changes into
the true ground state after some large number of iterations; however
the situation depends on the value of the rotational frequency
$\omega$.  In the figure, we see that the rotational frequencies
$\omega=0.5$, 1.0, and 1.5 MeV/$\hbar$ (the dashed, dotted, and
dot-dashed lines, respectively) lead to the true ground state. The
corresponding density distribution at the iteration step of 15000 is
plotted in Fig.~\ref{den1} (b).  The frequency $\omega=0.0$
MeV/$\hbar$ (the solid line) leads to the quasi-stable one (see
Fig.~\ref{den1} (c)).  At around $\omega=2.0$ MeV/$\hbar$, we obtain
the state (the thin solid line) with the $4\alpha$ linear chain
configuration, as shown in Fig.~\ref{den1} (d), whereas fission occurs
above those rotational frequencies (the dot-dot-dashed line).

We next estimate the range of the rotational frequencies where the
$4\alpha$ linear-chain configuration can be stabilized.
Figure~\ref{iner1} shows the coefficient of the rotational energy,
$\hbar^2/2\Theta$, versus the rotational frequency $\omega$ 
with various Skyrme interactions.
We find stable states for the $4\alpha$ linear chain
configuration.
For the SkI3 interaction, we obtain the
lower and upper bounds of the rotational frequencies
as 1.8 and 2.2 MeV/$\hbar$,
between these the
$4\alpha$ linear chain configuration is stabilized.
The values are
1.9 and 2.2 MeV/$\hbar$ for the SkI4 interaction 
and 2.0 and 2.1 MeV/$\hbar$ for the SLy6
interaction, respectively.  
In these frequency regions where the linear chain configuration is
stabilized, we can define the rigid-body moments of inertia, which is
calculated as 0.065 MeV for the SkI3 and SkI4 interactions and 0.06 MeV
for the SLy6 interaction.  There values are in very good agreement with
the conventional cluster model calculations\cite{Ita-O}.
  
We also estimate the corresponding angular momentum where the $4\alpha$
linear chain configuration is stabilized. We calculate the angular
momentum using the obtained rigid-body moment of inertia and compare
those with that calculated by the cranking method. The angular momentum
with the rigid-body moment of inertia, $J_{\rm rid}$, is calculated as
$J_{\rm rid}=\Theta \omega$.
The angular moment calculated using the cranking method, $J_{\rm cra}$,
is given by $J_{\rm cra}=<J>$, where $<J>$ is the expectation value of the
angular momentum in the cranking equation.
Figure~\ref{mom1} shows the angular
momentum thus obtained versus the rotational frequency. 
In the figure, we see that the calculated
angular momentum using the rigid-body moment of inertia agrees well with
that of the cranking method, indicating that the rigid-body
approximation is reasonable for the $4\alpha$ linear chain states.  We
find that the lower and upper bounds of the angular momentum where the
$4\alpha$ linear chain configuration is stabilized are about 13 and 18
$\hbar$ for the SkI3 interaction, 14 and 18 $\hbar$ for the SkI4
interaction, and 16 and 18 $\hbar$ for the SLy6 interaction,
respectively.  With such high angular momenta, 
very exotic configuration of $4\alpha$ linear chain can be stabilized,
however, fission occurs beyond this angular momentum region.

We analyzed such stability of linear chain configurations of $\alpha$ clusters
for three-$\alpha$ in $^{12}$C and three-$\alpha$ with valence neutrons
in $^{20}$C. However it is rather difficult to obtain stable regions
of rotational frequency and angular momentum in these cases, and
stability shown in $^{16}$O is considered to be a unique character of the
$4\alpha$ configuration. 

\subsection{Four-$\alpha$  linear chain in collisions of two $^8$Be with TDHF}

\begin{figure*}[t]
\begin{center}
\includegraphics[width=0.6\linewidth]{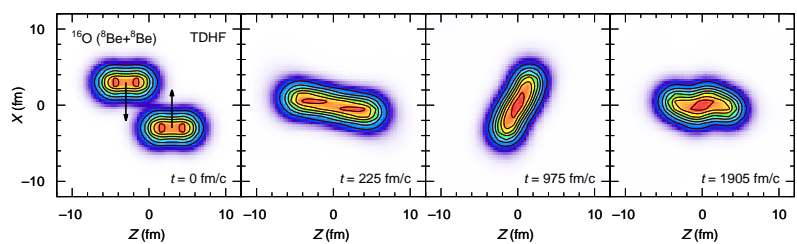}\\%
\end{center}
\caption{\label{tdhf1} Snapshots of the total nucleon
  density calculated using the TDHF method for the $^8$Be+$^8$Be
  collision. The impact parameter is 6 fm and the relative energy is 2
  MeV.  We normalize the color to the density distribution at the
  maximum of each plot.}
\end{figure*}

\begin{figure}[h]
\begin{center}
\includegraphics[width=0.6\linewidth]{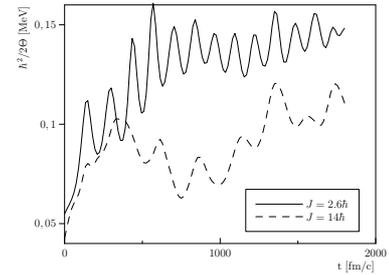}\\%
\end{center}
\caption{\label{tdhf2} Coefficient of the rotational energy,
 $\hbar^2/2\Theta$ calculated using the TDHF method for two
 situations: $E=3.3\,$MeV with $b=8\,$fm, corresponding to $J=2.6\hbar$,
 and $E=18\,$MeV  for the same impact parameter with $J=14\hbar$.}
\end{figure}

To check the importance
of this chain configuration as an intermediate state in a collision
situation, we also performed time-dependent Hartree-Fock calculations
for the $^8$Be+$^8$Be reaction in a configuration where the axes of
the collision partners are aligned. Fig.~\ref{tdhf1} shows a number of
typical snapshots of the collision for an impact parameter of 6~fm and a
relative energy of 2~MeV. The first picture shows the starting
configuration; the initial velocities are in the $x$-direction. Upon
contact the two nuclei are strongly attracted and form a compound
system that rotates while also undergoing strong vibrations as
indicated by the other snapshots. 

To illustrate the behavior of the moment of inertia we show its time
development for two typical cases in Fig.~\ref{tdhf2}. The moment of
inertia is calculated from the instantaneous density as the rigid-body
value.
They correspond
to two different angular momenta at the same impact
parameter of 8~fm, which is selected to get the system to coalesce
into a strongly deformed rotating configuration; the curves also show
the quite large oscillations caused by the shape changes. Yet the
tendency appears to follow that of the cranked calculations: for the
small angular momentum it goes to much smaller moment of inertia,
while for $J=14\hbar$ the system spends quite a long time oscillating
around the moment of inertia found in the cranking calculations before
relaxing to a slightly smaller value.

\section{Summary}
In this work we have performed microscopic dynamical calculations of
nuclear collisions to study the formation of a metastable linear chain
state of $^{12}$C.  The time evolution as calculated using the TDHF
equations shows a characteristic quasiperiodic exchange of alpha-like
clusters in the density function corresponding to a quasiperiodic
motion along a static Hartree-Fock potential energy surface, which is
studied using the density constraint procedure.  We have shown that
the calculations lead to the formation of a metastable linear chain
state of three alpha-like clusters which subsequently decays to a
lower energy triangular alpha-like configuration before acquiring a
more compact shape.  This is the first observation of such mode
changes in TDHF calculations and the results seem to be analogous to
the suggested astrophysical mechanism for the formation of $^{12}$C
nuclei.

For the $3\alpha$-chain configuration of $^{16}$O we found a
stabilization due to centrifugal forces for a range of angular
momenta not including zero. This is a new situation in cranked
mean-field calculations. An investigation of the presence of a chain
structure in collisions of $^8$Be+$^8$Be shows a rotating compound
system of roughly the right deformation, but with superimposed strong shape
oscillations.

\section*{Acknowledgments}
This work has been supported by the U.S. Department of Energy under
grant No.  DE-FG02-96ER40963 with Vanderbilt University, by the German
BMBF under contract No. 06FY9086, and by the GCOE program ``The Next
Generation of Physics, Spun from Universality and Emergence'' from
MEXT of Japan. A.S.U. thanks for support by the Hessian LOEWE
initiative through the Helmholtz International Center for FAIR; J.A.M
thanks the Japan Society for the Promotion of Science (JSPS) for an
invitation fellowship for research in Japan (short-term).

\end{document}